\pdfoutput=1
\documentclass[journal=jacsat,manuscript=article]{achemso}
\SectionNumbersOn

\usepackage[version=3]{mhchem} 
\usepackage{color}
\usepackage{graphics}
\usepackage{natmove}
\usepackage{bm}
\usepackage{float}
\setkeys{acs}{articletitle = true}
\author{Shih-Yang Lin}
\affiliation{Department of Physics, National Chung Cheng University, Chiayi 621, Taiwan}
\author{Wei-Bang Li}
\affiliation{Department of Physics, National Cheng Kung University, Tainan 701, Taiwan}
\author{Ngoc Thanh Thuy Tran}
\affiliation{Hi-Gem, National Cheng Kung University, Tainan 701, Taiwan}
\author{Wen-Dung Hsu}
\affiliation{Department of Materials Science and Engineering, National Cheng Kung University, Tainan 701, Taiwan}
\author{Hsin-Yi Liu}
\affiliation{Department of Physics, National Cheng Kung University, Tainan 701, Taiwan}
\author{Ming Fa-Lin}
\affiliation{Hi-Gem, National Cheng Kung University, Tainan 701, Taiwan}
\alsoaffiliation{QTC, National Cheng Kung University, Tainan 701, Taiwan}
\email{mflin@mail.ncku.edu.tw}
\title{Essential properties of Li/Li$^+$ graphite intercalation compounds}

\begin{document}


\begin{abstract}
The essential properties of graphite-based 3D systems are thoroughly investigated by the first-principles method. Such materials cover a simple hexagonal graphite, a Bernal graphite, and the stage-1 to stage-4 Li/Li$^+$ graphite intercalation compounds. The delicate calculations and the detailed analyses are done for their optimal stacking configurations, bong lengths, interlayer distances, free electron $\&$ hole densities, Fermi levels, transferred charges in chemical bondings, atom- or ion-dominated energy bands, spatial charge distributions and the significant variations after intercalation, Li-/Li$^+$- $\&$ C-orbital-decomposed DOSs. The above-mentioned physical quantities are sufficient in determining the critical orbital hybridizations responsible for the unusual fundamental properties. How to dramatically alter the low-lying electronic structures by modulating the quest-atom/quest-ion concentration is one of focuses, e.g., the drastic changes on the Fermi level, band widths, and number of energy bands. The theoretical predictions on the stage-n-dependent band structures could be examined by the high-resolution angle-resolved photoemission spectroscopy (ARPES). Most important, the low-energy DOSs near the Fermi might provide the reliable data for estimating the free carrier density due to the interlayer atomic interactions or the quest-atom/quest-ion intercalation. The van Hove singularities, which mainly arise from the critical points in energy-wave-vector space, could be directly examined by the experimental measurements of scanning tunneling spectroscopy (STS). Their features should be very useful in distinguishing the important differences among the stage-$n$ graphite intercalation compounds, and the distinct effects due to the atom or ion decoration.
\end{abstract}


\newpage
It is well known that graphite is one of the most investigated materials both theoretically and experimentally [Lin048] Up to now, it serves as the best anode in the Li$^+$-band battery. This layered system is purely composed of the hexagonal-symmetry carbon layers, in which the weak, but significant Van der Walls interactions would greatly modify the low-lying $\pi$-electronic structure and thus dominate the essential physical properties. A monolayer graphene is identified to be a zero-gap semiconductor [Refs], while a 3D graphite belongs to a semimetal, The electronic properties strongly depend on the way the graphitic planes are stacked on each other. In general, there are three kinds of stacking configurations: AAA (simple hexagonal, ABAB (Bernal), and ABCABC (rhombohedral). The total free carrier density is predicted to be ${3.5\times\,10^{20}}$ e/cm$^3$ in simple hexagonal graphite [Res] and ${\sim\,10^{19}}$ in Bernal graphite at room temperature [Refs]. When various atoms and molecules are further intercalated into into the AB-stacked graphite, a lot of graphite intercalation compounds are formed [Refs]. When many free conductions (holes) are induced after the intercalation, such systems exhibit the donor-type (acceptor-type) behaviors. Among these compounds, only the stage-$n$ lithium intercalation systems display the AAA stacking configuration, as confirmed from the X-ray diffraction patterns [Refs]. Here $n$ clearly indicates the number of graphitic sheets between two periodical quest-atom layers, in which n=1, 2, 3 and 4 (Figs. 1(c)-1(f)), being arranged from the highest concentration to the lower one, will be studied thoroughly in terms of the essential properties. It is also noticed that the other alkali intercalation compounds present the MC$_8$ structure in the stage-1 configuration (M=K, Rb; Cs) [Refs]. Obviously, both
LiC$_6$ and  MC$_8$ should have very different band structures and fundamentla properties.

Up to now, there are a lot of theoretical and experimental researches on alkali-atom graphite intercalation compounds [Refs], but only few studies about alkali-ion ones [Refs]. As to the former, the first-principles method\cite{toyoura2008first,persson2010thermodynamic,toyoura2010effects,wang2014van} has been utilized to investigate the total ground state energies, optimal guest-atom distributions, interlayer distances, stacking configuration between neighboring graphitic layers, intrinsic and atom-doped electronic energy spectra, and density of states, transport properties, and phonon spectra Such systems cover LiC$_x$ ($x$=6, 12, 18; 24) [Refs], and MC$_x$ ($x$=8, 16, 24; 32) [Refs]. Specifically, the thermodynamic and kinetic properties of lithium atoms in graphite intercalation compounds are thoroughly explored by the first-principles calculations\cite{toyoura2008first,persson2010thermodynamic,toyoura2010effects} and Monte Carlo method.\cite{persson2010thermodynamic} The Li-atom diffusions in graphite are classified into three types: (I) vacancy, (II) interstitial, and (III) interstitialcy mechanisms. The similar numerical investigations are applied to the Li$^+$-ion diffusions on the graphene layers,\cite{zheng2011diffusion,li2018ultra} in which their results cover the optimal height and site of guest atoms, the absorption energy, and energy barrier along the specific transport path. Moreover, the tight-bind model and the superlattice model are frequently utilized to study the fundamental properties for pristine graphites and the atom/molecule-intercalated graphite compounds, e.g., optical absorption spectra [Refs], Coulomb excitations [Refs], and magnetic properties [Refs]. As to the various expeirmental measurements on graphite-related systems, they include the X-ray/TEM/STM/LEED patterns [Refs], angle -resolved photoemission spectra [Refs], optical reflectance/abosprtion/transmission spectra [Refs], electrical conductivities [Refs], magnetic properties [Refs], phonon energy spectra [Refs], electron energy loss spectra [Refs], and femtosecond excited carrier dynamics [Refs]. Specifically, the planar Li-atom distribution configurations of stage-1, stage-2 and stage-3 compounds, as well as their periodical interlayer distances, are examined by the X-ray diffraction spectra [Refs]. The intercalation-induced high conduction-electron densities are also observed in the optical [Refs] and transport measurements [Refs]. The rich and unique phenomena in alkali-atom graphite intercalation compounds are expected to have the significant differences under a systematic comparison with those in alkali-ion cases.

\vspace{5mm}
\par\noindent
{\bf The theoretical model}

A typical condensed-matter system is made of a periodic crystal potential, where each atom would contribute several valence electrons around an ionic core. Obviously, each one exhibits the complex composite effects mainly arising the electron-electron Coulomb interactions and the electron-ion crystal potential, especially for the many-body effects. This becomes a high barrier in solving the many-particle Schrodinger equation. The difficulties of numerical calculations are greatly enhanced when the various chemical environments need to be taken into consideration,. e.g., the very weak van der Waals interactions between two neighboring graphitic layers and the chemisorptions on graphite surface [Refs]. Some approximate methods have been proposed to achieve the reliable geometric structures and electronic properties. Up to date, the first-principles calculations are frequently utilized to obtain the quantum states of periodic systems. Such numerical calculations are very efficient for fully exploring the fundamental physical properties. Specifically, Vienna ab initio simulation package (VASP)\cite{kresse1996efficient} evaluates an approximate solution within the density functional theory by solving the so-called Kohn-Sham equations.\cite{kohn1965self} The charge distribution can determine all the intrinsic interactions in a condensed-matter systems; that is, the carrier density is responsible for the ground state energy and the essential properties. The spatial charge density could be solved by the numerical self-consistent scheme, as clearly shown in a flow chart of detailed evaluations [Fig. 1]. Compared with the first-principles method, the tight-binding model, with the hopping integrals (the parameters), cannot study the optimal geometric structures and thus identify the complicated orbital hybridizations in various chemical bonds. By a detailed comparison of these two modes in the low-lying energy bands, the reliable parameters of the latter are thus obtained, e.g., the vertical and non-vertical atomic interactions in AA- [Refs], AB- [Refs] and ABC-stacked graphites [Refs]. And then, they are very useful in understanding other essential properties, e.g., optical properties [Refs] and Coulomb excitations [Refs] in three kinds of graphites. In addition, the generalized tight-binding model, but not the first-principles method, is suitable for studying the essential properties under the external fields, e.g., the quantized Landau levels in a uniform magnetic field [Refs]

In this Chapter, the optimal geometric structures and electronic properties and magnetic configurations are thoroughly studied for 3D graphite-related systems by utilizing VASP.\cite{kresse1996efficient} Such systems cover monolayer graphene, Bernal graphite, simple hexagonal graphite, and stage-1 to stage-4 Li/Li$^+$ graphite intercalation compounds. The Perdew-Burke-Ernzerhof functiona within the generalized gradient approximation can deal with the many-particle Coulomb interactions, the exchange and correlation energies of valence and conduction electrons.\cite{perdew1996generalized} The projector-augmented wave pseudo-potentials characterize the electron-ion interactions.\cite{blochl1994projector} It should be noticed that to correctly describe the weak but significant atomic interactions between the neighboring graphitic layers, the van der Waals forceis must be included in the calculations by the semi-empirical DFT-D2 correction of Grimme.\cite{grimme2006semiempirical} When one solves the many-body Schrodinger equation, plane waves, with an maximum energy cutoff of 400 eV, consists of a complete set in building the Bloch wave functions.
For charged systems calculations, monopole, dipole and quadrupole corrections in all directions are calculated. The 3D periodic boundary condition is along ${\hat x}$, ${\hat y}$, and ${\hat z}$, in which a primitive unit cell depends on the geometric distribution of host and guest atoms. The Brillouin zone is sampled by ${30\times\,30\times\,30}$ and ${70\times\,70\times\,70}$ ${\bf k}$ point meshes within the Monkhorst-Pack scheme, respectively, corresponding to the numerical calculations of geometric optimizations and electronic structures. The energy convergence is set to be ${10^{-5}}$ eV for two neighboring simulation steps; furthermore, the maximum Hellmann-Feynman force acting on each atom is less than 0.01 eV/$\AA$ during the process of ionic relaxations. The details of the optimal calculation processes are revealed in Fig. 1.

\vspace{5mm}
\par\noindent
{\bf Rich geometric structures of graphites and graphite intercalation compounds}

Apparently, the pristine graphites and Li/Li$^+$ graphite intercalation compounds display the rich geometric structures by the VASP calculations on the total ground state energies. The periodical graphitic layers in the AA stacking are bound together by the weak van der Waals interactions due to the perpendicular ${2p_z}$ orbitals on two neighboring planes, so it needs to have the delicate numerical evaluations on the geometry-dependent total ground state energies. These intrinsic interactions can create the vertical and non-vertical hopping integrals in the tight-binding model and thus dominate the low-energy electronic properties [Refs]. Of course, the very strong $\sigma$ bondings also exist on each graphitic layer. They are only slightly changed during the varation from a 2D graphene to a 3D graphite, as confirmed by the C-C bond lengths of them (1.420 $\AA$ and 1.423 $\AA$, respectively). By the detailed calculations in Fig. 2, a simple hexagonal graphite is identified to present a optimal interlayer distance of ${I_z=3.489}$ $\AA$, with the lowest ground state energy of ${-0.238}$ eV. This system possesses the largest interlayer distance and the highest ground state energy, compared with those [(3.35 $\AA$, ${-0.6}$ eV) and (3.40 $\AA$, ${-0.4}$ eV) of Bernal graphite and rhombohedral graphite. Such significant differences are responsible for the fact that a natural graphite is made up of AB and ABC stackings, especially for the former configuration. The distinct stacking symmetries would dominate the interlayer atomic interactions and greatly diversify the fundamental chemical and physical properties.

The geometric symmetries are greatly diversified by the chemical intercalation. The Li/Li$^{+}$ could be easily intercalated into graphitic layers, so that such guest atoms/ions create a periodical planar structure. According to their concentrations, there exist the stage-dependent graphite intercalation compounds, being clearly identified in the experimental syntheses.\cite{guerard1975intercalation,kambe1979intercalate,basu1979synthesis,dicenzo1981plane,billaud1996revisited} Three types of absorption positions, hollow, top and bridge ones, are frequently observed in the engineering of chemical modification.\cite{chan2008first} Among them, the hollow-site positions are examined to have the lowest ground states, as clearly indicated in Table 1. That is to say, such positions present the strongest chemical bondings/interactions. Similar results are revealed in other alkali-atom/ion intercalations.\cite{dresselhaus2002intercalation,okamoto2013density} By the detailed calculations, the height-dependent (interlayer-distance-dependent) ground state energies show the optimal interlayer distance with the lowest one. For example, the stage-1 LiC$_6$/Li$^+$C$_6$ has the interlayer distance of 3.728 $\AA$/3.059 $\AA$ under the ground state energy of ${-58.689}$ eV/${-65.417}$ eV. Obviously, the ground state energy decreases as the stage number of $n$ grows, in which the declining concentration of Li-C/Li$^+$-C bonds is the main reason [Check for  the ion cases]. Compared with the original interlayer distance of two neighboring graphitic layers 3.489 $\AA$ in Fig. 2(a)], those with and without the Li/Li$^+$ intercalations are, respectively, enhanced and reduced except for the Li$^+$C$_6$ case, e.g., ${\sim\,3.728-3.768}$ $\AA$/3.059$-$3.539 $\AA$ and ${\sim\,3.323-3.364}$ $\AA$/2.918$-$3.348 $\AA$ under the stage-1 to stage-4 configurations. A simple relation between $n$ and the interlayer distances might be absent. However, a detailed comparison between the Li and Li$^+$ intercalation cases clearly show that the latter always have the shorter interlayer distances corresponding to the lower total ground state energies. Apparently, such significant results suggest the stronger intrinsic interactions in Li$^+$ intercalation compounds and more stable ion status; that is, they might account for the rapid charging and discharging closely related to the intercalation and deintercalation processes, respectively {REfs]. As to the C-C bond lengths, they are only slightly changed by modified by the Li-atom and Li$^+$-ion intercalations. The stage-1, stage-2, stage-3 and stage-4 Li (Li$^+$) compounds, respectively, have 1.443 $\AA$, 1.429 $\AA$, 1.428 $\AA$, 1.426  $\AA$ (1.414 $\AA$, 1.414 $\AA$, 1.416 $\AA$ and 1.418 $\AA$). Their enhancement and decrease in the atomic and ionic cases, suggesting charge transfers of ${Li\rightarrow\,C}$ and ${C\rightarrow\,Li{^+}}$. This is consistent with the calculations of Bader analyses [Table 2].

It is worthy of a closer examination about charge transfers ($f$'s) stacking configurations (AA, AB, ABC). By the Bader analyses in the VASP calclations, the stage-1, stage-2, stage-3 and stage-4 Li-intercalated (Li$^+$-based) graphite compounds, respectively, exhibit the charge transfers of 0.855, 0.857, 0.873 and 0.876 (0.183, 0.138, 0.117, and 0.132), being almost independent of the number of stage $n$. The ${2s}$ orbitals of Li atoms are mostly transferred to the neighboring C atoms, while they do not present the 100$\%$ transfer. The 3D free electron density is deduced to be inversely proportional to $n$; that is, $f$ could be utilized to evaluate the conduction electron density. Also, it could be estimated from the area covered by the curve of density-of-state versus energy (discussed later). The ${2s}$ orbitals of Li atoms are mostly transferred to the neighboring C atoms, while they do not present the 100$\%$ transfer. To easily compare all the calculate results, the stacking configuration is assumed to be the sequence of AA. However, the very low guest-atom/ion concentration or the pristine case could belong to the AB stacking, as observed in a natural graphite. How to transform from the AA to AB stackings during the decline of guest atoms or ions is an interesting focus of the future studies.

Scanning tunneling microscopy (STM) is the most powerful experimental technique in resolving the surface structure, being  able to characterize the surface topographies in real space with both lateral and vertical atomic resolution, e.g., the nano-scaled bond lengths, crystal orientations, planar/non-planar structures, step edges, local vacancies, dislocations, atomic or molecular adsorptions, and nanoclusters. The up-to-date STM measurements on the graphite-related systems have confirmed the complex relations among the hexagonal lattice, the finite-size confinement, the flexible feature, and the diverse chemical bondings of carbon atoms. For example, graphene nanoribbons exhibit the nanoscale-width planar honeycomb lattice, accompanied with the achiral (armchair and zigzag) or chiral edge structures.\cite{ruffieux2012electronic,tao2011spatially,magda2014room} Furthermore, they could also be formed in the curved,\cite{tao2011spatially} folded,\cite{li2008chemically} scrolled\cite{viculis2003chemical} and stacked lattice structures.\cite{johnson2010hydrogen} Carbon nanotubes possess the achiral or chiral arrangements of hexagons on a cylindrical surface.\cite{wilder1998electronic,odom1998atomic} The atomic-scale measurements directly identify the AB,\cite{xu2012new,xu2012electronic} ABC\cite{yankowitz2014electric,lauffer2008atomic} and AAB\cite{rong1993electronic} stacking configurations in few-layer graphene systems, the corrugated substrate and buffer graphene layer,\cite{chen2015long,chen2015tailoring} the rippled structures of graphene islands,\cite{meng2013strain,bai2014creating,de2008periodically} the adatom distributions on graphene surfaces.\cite{pandey2008scanning,balog2010bandgap}The layered graphite could exhibit the 2D networks of local defects on surface; the pyridinic-nitrogen and graphitic-N structures.\cite{kondo2012atomic} Apparently, the high-resolution experimental measurements of STM could be utilized to verify the unique surface structures of Li-atom and Li$^+$-ion graphite intercalation compounds, e.g., the C-C bonds affected by the guest-atom or guest-ion intercalation, the optimal adsorption site, and the planar periodical distribution of host atoms and guest atoms/ions.

Transmission electron microscopy is a microscopy technique in which an electron beam, with a uniform current density, is transmitted through an ultra thin specimen to create an image, as a result of the interactions between incident charges and sample. TEM is a very important experimental technique in directly visualizing the crystal structure, locating and identifying the type of defects, and studying structural phase transitions. This measurement  has a rather big electrons' atomic scattering factor, being $\sim$10000 times of that from the X-ray diffraction. TEM provides electron diffraction an advantage to observe even the weakest diffracted spot. However, its resolution is seriously limited by spherical and chromatic aberrations of the lenses. More delicate techniques for improving the diffraction resolution become indispensable. By applying a monochromator and a Cs corrector into TEM, which is called the high-resolution TEM (HRTEM), the structural resolution can reach less than 0.5 {\AA}. HRTEM has been successfully and extensively utilized to analyze crystal structures and lattice imperfections in various nanomaterials. The TEM/HRTEM measurements on graphene-related systems are very suitable in identifying the ${sp^2}$-bonding-enriched nanoscale structures, such as the multi-walled cylindrical structures of carbon nanotubes,\cite{iijima1991helical,kosynkin2009longitudinal} the curved,\cite{kosynkin2009longitudinal,cataldo2010graphene,kumar2011laser} folded\cite{liu2009open,zhang2010free} and scrolled profiles\cite{viculis2003chemical,savoskin2007carbon,shioyama2003new} of graphene nanoribbons, as well as the stacking configurations and the interlayer distances of few-layer graphene systems.\cite{campos2008bulk,warner2008direct,lee2008growth} Obviously, the high-resolution TEM measurements are suitable in examining the rich geometric properties of pristine (AA and ABC) and Li- $\&$Li$^+$-intercalated graphites, such as, the periodical distance along the $z$-direction, the distances of layered graphene in the high-$n$ systems the stacking configuration of graphitic layers, and the optimal distribution configuration $\&$ position of guest atoms or ions.

It is well known that X-ray diffraction techniques could provide the most powerful tools in exploring the crystal symmetries, especially for the 3D condensed systems.\cite{guerard1975intercalation,kambe1979intercalate,basu1979synthesis,dicenzo1981plane,billaud1996revisited} Specifically, Such experimental measurements are very suitable for the stage$-n$ structures in Li-/Li$^+$-graphite intercalation compounds, e.g., the lattice constants, interlayer distances and order/disorder transformations. Up to now, the X-ray diffraction patterns have been utilized to confirmed the existence of stage-1,\cite{guerard1975intercalation,kambe1979intercalate,basu1979synthesis,billaud1996revisited}  stage-2,\cite{guerard1975intercalation,basu1979synthesis,billaud1996revisited}  and stage-3\cite{dicenzo1981plane} Li-graphite intercalation compounds. The optimal planar structure belongs to ${\sqrt 3\times\,\sqrt 3\,R30^\circ}$ with a lattice constant ${\sim\,4.305}$ $\AA$, as observed from both LiC$_6$ and LiC$_{12}$.\cite{guerard1975intercalation,basu1979synthesis,billaud1996revisited} The periodical interlayer distances in the stange-1 and stage-2 are, respectively, identified to be ${\sim\,3.705-3.706}$ $\AA$ and ${\sim\,7.025-7.065}$ $\AA$. Furthermore, the graphitic layers of them present the AA stacking along the $z$-direction. However, the stacking ordering of ABA$\alpha$ABA$\alpha$..., is examined to survive in the stage-3 system (LiC$_{18}$).\cite{dicenzo1981plane} The high-resolution X-ray measurements are required to thoroughly examine the stacking configurations of the larger-$n$ Li-graphite intercalation compounds. The similarities and important differences between the stage-$n$ Li-atom and Li$^+$-ion systems in the optimal geometric structures are worthy of the detailed examinations using the X-ray diffraction spectra, especially for the experimental measurements on the latter. They could reveal  the important informations about the available electronic  configurations associated with the neutral atoms and full ions in the anode of Li$^+$-based battery.

\vspace{5mm}
\par\noindent
{\bf Unusual band structures of graphite-related systems}

A pristine graphite possesses an unusual electronic structure with a large or small overlap of valence and conduction bands, being sensitive to the stacking configuration, e.g., the AA, AB and ABC stackings associated with Figs. 4(a), 4(b) and 4(c), respectively. It is well known that a monolayer graphene is a zero-gap semiconductor.\cite{geim2010rise} There exist the linearly intersecting valence and conduction bands at the Dirac point (not shown), so the density of states at the Fermi level (the band overlap) is vanishing. Only the finite temperature can induce some free carrier density.\cite{bolotin2008temperature} The low-energy electronic states are initiated from two equivalent valleys. However, the significant interlayer atomic interactions lead to the drastic changes of the low-lying energy bands. All the 3D graphitic systems, the occupied valence bands are asymmetric to the unoccupied conduction bands about the Fermi level, mainly owing the significant interlayer hopping integrals. Among three kinds of graphites, the AA-stacked graphite has a strong vertical interlayer hopping integral.\cite{charlier1992first} The Dirac-point energies, corresponding to the corner states in the first Brillouin zone [Fig. 2(c)], strongly depends on the $k_z$-components of 3D wave vectors. As a result, an obvious band overlap is revealed in the $k_z$-dependent energy dispersion along $\Gamma$A. The periodical arrangement of graphitic layers in the $z$-direction can create the band width as wide as ${\sim\,1}$ eV. Furthermore, on the ${(k_x,k_y)}$-plane, an electron (hole) pocket comes to exist near the A ($\Gamma$) point. That is to say, the free electrons and holes, which appears in the density of states [Fig. 7(a)], are, respectively, located in the range of ${0\le\,E\le\,0.5}$ eV and ${-0.5}$ eV${\le\,E\le\,0}$. The above-mentioned low-lying energy bands, with ${|E^{c,v}|\le\,3}$ eV, mainly come from the $\pi$ bondings of the C-2p$_z$ orbitals. Such electronic states are responsible for most of the essential physical properties. Specifically, the parabolic energy dispersions belong to the saddle points at the the middle points of the first Brillouin zone, e.g., the electronic energy spectra at the M and L points [2.5 eV and 1 eV] $\&$ [$-$2 eV and $-$3 eV], as shown in Figs. 4(a) and 4(d). They possess very high density of states [discussed later in Figs. 8-10], so their features are useful in distinguishing the diversified intercalation effects. As to the deeper-energy electronic states (${E^v \le -3}$ eV), part of them originates the $\sigma$ bondings of the ${(2p_x,2p_y,2s)}$ orbitals, especially for those near the $\Gamma$ and A points (${\sim\,-3}$ eV). In addition, the initial $\sigma$-electronic states just has the same energy with the $\pi$-electronic saddle point of L. They will exhibit the splitting behavior after intercalation.

There are certain important differences among there typical kinds of graphites. The linear Dirac-cone structures on the ${(k_x,k_y)}$ planes are clearly revealed in the AA- [Fig. 3(a)] and ABC-stacked graphites [Fig. 3(c). For a simple hexagonal graphite, such unusual energy bands are just initiated from the corners of the original/reduced first Brillouin zone. Furthermore, a rhombohedral graphite exhibits a novel spiral configuration along the $k_z$-direction (a slight deviation from the planar coner), mainly owing to the specific stacking symmetry.\cite{ho2013precessing} In general, the energy of Dirac points are very sensitive to the ${k_z}$ component of wave vector. On the other side, a Bernal graphite presents the bilayer- and monolayer-like energy dispersions (the parabolic and linear ones), as indicated in Fig. 3(b). This result is consistent with the previous theoretical calculations and the experimental ARPES measurements. In term of the $k_z$-dependent $\pi$-electronic band widths related to the ${2p_z}$ orbitals [Figs. 3(a)-3(c)], they are, respectively, about 1.0 eV, 0.2 eV and 0.03 eV for simple hexagonal, Bernal and rhombohedral graphites. such as, those near the Fermi level along KH and at the middle energy along ML  (${\pm\,2}$ eV). Obviously, the first and third systems have the largest and smallest valence and conduction overlaps, respectively, and so do the free electron/hole density purely due to the interlayer atomic interactions. The above-mentioned  results suggest that the interlayer hopping integrals of the tight-binding model [REFs] are most complicated for the ABC-stacked graphite, when the critical parameters are fitted from the first-principles calculations [REfs]. In short, the AA-, AB- and ABC-stacked graphites belong to semimetals, while the latter two only possess very few 3D free carrier densities at low temperatures. These systems are thus expected to display the diverse physical and chemical properties, e.g., the unusual magnetic quantization phenomena [REfs],  magneto-optical properties [REfs], and many-particle Coulomb excitations[Refs].

Electronic structures exhibit the drastic changes after the intercalation of Li atoms, as clearly illustrated in Figs. 5(a)-5(d) for stage-1 to stage-4 graphite intercalation compounds, respectively. For the stage-1 LiC$_6$, the asymmetry of electron and hole bands, being about the Fermi level, become more serious, as identified from a detailed comparison with the pristine case [Figs. 4(d) and 5(a)]. Apparently, the main mechanisms originate from more orbital hybridizations due to the Li-C bonds [Fig. 3(c)]. Apparently, the Fermi level presents the blue-shift phenomenon, in which it is situated at certain conduction bands and does not intersect with any valence bands. The free conduction electrons are purely induced by the Li-atom intercalation; furthermore, the free holes related to the unoccupied valence states thoroughly disappear. That is, two kinds of free carriers arising from the interlayer atomic interactions are fully replaced by the conduction ones. It should be noticed that the energy bands just below the Fermi level are split along the $\Gamma$A direction, and the originate double degeneracy, as shown in Fig. 4(d), is only created by the zone-folding effect with an enlagred unit cell with six carbon atoms. The upper and lower energy bands, respectively, represent the conduction and valence states. The maximum and minimum values of the valence band (indicated by the red dashed curves) are ${-0.6}$ eV and ${-2.01}$ eV ($\Gamma$ and A points). The latter, the maximum energy range between the Fermi level and the top of valence state,  plays a critical role in determining the blue-shift value of the Fermi level, i.e., it will determine the Li-intercalation-induced 3D conduction electron density. Moreover, the low-lying energy bands do not present the crossing and anti-crossing behaviors between any two energy subbands. Specifically, the $\pi$-electronic energy spectra at the saddle points of L and M are not identical to the initial $\sigma$-electronic ones along $\Gamma$A. This will lead to the separation of van Hove singularities.

Band structures are getting more complex during as the number of stage $n$ grows. Concerning stage-2 to stage-4 compounds [Figs. 5(b)-5(d)], there are more energy subbands as a result of the enlarge unit cell. However, the scale of the first Brillouin zone is greatly reduced. The crossing and anti-crossing behaviors come to exist frequently.  Obviously, the effective blue shift of the Fermi level, the maximum energy range of $E_F$ and the top valence state at the A point, which is investigated later by a more accurate method for density of states [Figs. 9(a)-9(d)], presents the declining behaviors. The stage-2 LiC$_{12}$ has two valence subbands slightly crossing the Fermi level near the $\Gamma$ point [Fig. 5(b)]. The similar phenomenon becomes obvious and even appears at the A point for the LiC$_{18}$ and LiC$_{24}$ compounds [Figs. 5(c) and 5(d)]. Apparently, the 3D conduction electron density decreases in the increment of $n$. Moreover, the $\pi$-electronic parabolic dispersions, being near the  M and L, exhibit the obvious splitting behaviors. There exist (one, two, three, four) valence and conduction saddle points, respectively, in the stage-1, stage-2, stage-3 and stage-4 compounds. They represent the most important feature in distinguishing the diverse intercalation effects. The $n$-enriched electronic structures lie in the drastic changes of the quest-atom concentration and their distribution configuration.

The Li$^+$-ion intercalation can greatly diversify electronic structures, as examined from a detailed comparison with the pristine and  guest-atom cases. For example,  Li$^+$C$_6$, LiC$_6$ and AA-stacked graphite [Figs. 6(a), 5(a), and 4(d)] are rather different from one another in the electronic properties, especially for the low-lying valence and conduction bands. The stage-1 Li$^+$C$_6$ exhibits two splitting energy subbands along $\Gamma$A near the Fermi level. Furthermore, the conduction and valence ones are partially occupied and unoccupied, respectively, being clearly illustrated by the energy dispersions along AH and $\Gamma$M. However, the valence bands are fully occupied in the stage-1 LiC$_6$ [Fig. 5(a)]. Obviously, the splitting behavior indicates the separation and distortion of the linear Dirac-cone structure [Figs. 6(a) and 4(d)], e.g., the energy spacing of ${\sim\,0.45-0.9}$ eV in the separated valence and conduction bands.  The linearly gapless Dirac cones, being revealed in a simple hexagonal graphite, might be vanishing through  the Li$^+$ intercalation. According to various band structures near the Fermi level [Figs. 6(a)-6(d)], the stage-$n$ graphite Li$^+$-ion intercalation compounds are predicted to have the same free electron and hole densities, being similar to those in the AA-stacked graphite. That is, they do not possess the high-density conduction electrons, or the charge transfer almost disappears during the intercalation/deintercalation of guest Li$^+$ ions. The Fermi level almost remains at the middle between the unoccupied valence states and the occupied conduction ones. Moreover, energy bands become more complicated with the increasing stage $n$.

On the experimental side, ARPES is best for exploring the quasi-particle energies of the occupied electronic states within the Brillouin zone,\cite{hufner2013photoelectron} and the measured dispersion relations can directly examine those evaluated from the first-principles method [Refs] tight-bindig model [Refs]. The ARPES chamber is associated with instrument of sample synthesis to identify the in-situ band structures. Up to date, the measured results have confirmed the feature-rich energy bands in the carbon-related systems, as verified under the distinct dimensions,\cite{ohta2007interlayer,siegel2013charge,bostwick2007quasiparticle,ruffieux2012electronic,sugawara2006fermi,gruneis2008electron} layer numbers,\cite{coletti2013revealing,ohta2007interlayer,bostwick2014coexisting,sutter2009electronic} stacking configurations,\cite{ohta2006controlling} substrates,\cite{coletti2013revealing,ohta2007interlayer,siegel2013charge} and adatom/molecule chemisorptions.\cite{papagno2011large,zhou2008metal,gruneis2008tunable} Graphene nanoribbons are identified to exhibit 1D parabolic energy subbands centered at the high-symmetry point, accompanied with an energy gap and different subband spacings.\cite{ruffieux2012electronic} Recently, a plenty of ARPES measurements are done for few-layer graphenes, obviously showing the linear Dirac-cone structure of monolayer system,\cite{ohta2007interlayer,siegel2013charge,bostwick2007quasiparticle} two pairs of parabolic bands in bilayer AB stacking,\cite{ohta2007interlayer,ohta2006controlling} the co-existent linear and parabolic dispersions in symmetry-broken bilayer system,\cite{bostwick2014coexisting} the monolayer- and bilayer-like energy bands in tri-layer ABA stacking,\cite{coletti2013revealing,ohta2007interlayer} the linear, partially flat and sombrero-shaped energy bands of tri-layer ABC stacking,\cite{coletti2013revealing} the substrate-induced large energy spacing between the $\pi$ and $\pi^\ast$ bands in bilayer AB stacking,\cite{coletti2013revealing,ohta2007interlayer} the substrate-created oscillatory bands in few-layer ABC stacking,\cite{ohta2007interlayer} and the metal-semiconductor transition and the tunable low-lying energy bands after the adatom/molecule chemisorptions on graphene surface.\cite{papagno2011large,zhou2008metal,gruneis2008tunable} It should be noticed that the high-technique ARPES is required to examine the 3D band structure of the AB-stacked natural graphite,\cite{sugawara2006fermi,gruneis2008electron} since the conservation of the transferred momentum is destroyed along the $k_z$-direction [REFs]. The experimental measurements clearly illustrate that the 3D energy bands the bilayer- and monolayer-like energy dispersions at ${k_z=0}$ and 1 (the K and H in the first Brillouin zone), respectively; furthermore, there exists the strong trigonal warping effect around the KH axis, and a weakly dispersive band near the Fermi energy due to the stepped surface with zigzag edge.

The high-resolution ARPES is required to examine the theoretical predictions on the occupied valence and conduction states of graphite-relate systems, covering electronic energy spectra of simple hexagonal graphite [Fig. 4(a)], rhombohedral graphite [Fig. 4(c)], and stage$-n$ Li-atom [Figs. 5(a)-5(d)] $\&$ LI$^+$-ion graphite intercalation [Figs. 6(a)-6(d)]. Such experimental measurements could provide the full information on the important effects due to the stacking configuration, the significant Li-C chemical bonding (charge transfer), and the zone folding (the lower symmetry of Li$^+$-ion distribution).
They would identify the critical electronic structures below the Fermi level. The detailed eaminations, which are conducted on the ${k_z}$-dependent band widths and linear Dirac-cone structures, are sufficient in distinguishing the AA and ABC stacking cnfigurations.  The experimental identifications on the large blue shift of the Fermi level is an obvious evidence about the Li-atom intercalation effect (the strong charge transfer from Li to C). A declining relation between $E_F$ and stage $n$ is reflected in the red shift of the lower conduction subbands (the higher valence ones). However, the Fermi level is almost identical for the pristine and Li$^+$-intercalation cases. The latter exhibit the splitting of energy subbands along the $\Gamma$A direction, in which  this behavior become more serious in the increment of $n$ because of the reduced distribution symmetry. Another important focuses of Li/Li$^+$ intercalation effects cover the $n$-split $\pi$-electronic parabolic dispersions near the saddle points of M and L and their state energies separated from the initial $\sigma$-electronic ones.

\vspace{5mm}
\par\noindent
{\bf van Hove singularities in density of states}

The van Hove singularities in 3D graphite-related systems are revealed as the special structures, being quite different from those in layered graphene systems [Refs]. The atom- and orbital-decomposed density of states, as shown in Figs. 9, 10, and 11, could provide the full information and be very useful in examining the atom/ion intercalation effects. Their structures principally originate from the critical points in the energy-wave-vector space, in which such points might form a 1D space with a continue $k_z$-dependence. For a simple hexagonal graphite [Fig. 9(a)], density of states is asymmetric about the Fermi level because of the interlayer atomic interactions. It belongs to a semimetal, since DOS, a finite but minimum value at ${E=0}$, being situated in the ranges of ${0<E\le\,0.5}$ eV and ${-0.5}$ eV ${\le\,E<0}$ eV, respectively, is created by free electrons and holes [Fig. 4(d)]. Apparently, the dip structure centered at the Fermi level is closely related to the $k_z$-dependent Dirac-cone structures. A similar van Hove singularity is induced by the Dirac point in monolayer graphene [Refs]. At the deeper/higher energy, the prominent and wide shoulder structures come to exist, corresponding to the saddle points on the ${(k_x,k_y)}$-plane of the $\pi$-electronic energy bands, e.g., the M and L points in Fig. 4(d). Their energy ranges of valence and conduction states are, respectively, ${-3}$ eV ${\le\,E\le\,-2.0}$ eV and ${1.0}$ eV ${\le\,E<2.5}$ eV. As to the $\sigma$ orbitals of ${(2p_x,2p_y,2s)}$, the significant contributions only appear at ${E\le\,-3.0}$ eV and are are absent in the conduction region. This clearly illustrates the very strong $\sigma$ bondings in forming the planar geometric structure. Specially, the $\pi$- and $\sigma$-electronic spectra might have significant contributions at the same energy of ${E\sim\,-3}$ eV. The AA-stacked graphite is very different from the AB- and ABC-stacked ones [Figs. 9(b) and 9(c)] in terms of density of states at the Fermi level, the energy range of band overlap, and the widths of shoulder structures. Apparently, the first (third) system exhibits the most prominent (weak) features, mainly owing to the highest (lowest) geometric symmetry, or the strongest (smallest) interlayer hopping integrals.

The Li-atom intercalation has created the dramatic transformation of density of states, as obviously illustrated in Fig. 10(a)-10(d) for stage-1 to stage-4 graphite compounds. Compared with density of states of a simple hexagonal graphite [Fig. 9(a)], such Li-intercalation systems exhibit more asymmetric energy spectra about the Fermi level. These results clearly indicate the large blue shifts of the Fermi level, being determined by a certain negative energy with a minimum density of states (a green arrow) and $E_F$ (0 eV). They are estimated to be, respectively, 1.45 eV, 1,10 eV, 0.95 eV, and 0.80 eV for the ${n=1}$, 2, 3, and 4 systems. Specifically, the significant Li-C chemical bondings lead to the drastic changes of conduction and valence energy spectra, especially for those near the saddle points (e.g., the M and L points). The energy range and the number of the prominent shoulder/peak structures strongly depend on the stage $n$. For LiC$_6$ [Fig 10(a)], the widths of  valence and conduction shoulders are, respectively, widened and reduced, in which the former separates from that due to the $\sigma$ orbitals (the green and dashed black curves). As a result, the most important features in distinguishing the pristine and stage-1 cases are the largest blue shift of $E_F$ and the separated contributions of the $\pi$ and $\sigma$ orbitals (the three separated structures within $-$4.8 eV${le\,E\,-3.5}$ eV). With the increasing $n$, there exist two, three and four well-separated peak structures in the valence and conduction DOSs of stage-2, stage-3 $\&$ stage-4 compounds, respectively [Figs. 10(b), 10(c) and 10(d)]. Furthermore, one and two shoulder structures due to the initial contributions of the $\sigma$ orbitals appear at the deeper energy (${E\sim\,-.4.3}$ eV-${-3.2}$ eV), respectively, corresponding to  (stage-1 $\&$ stage-2) and (stage-3 $\&$ stage-4) systems. These results directly reflect the stage-dependent chemical environments experienced by the $\pi$ and $\sigma$ chemical bondings (perpendicular and parallel to graphitic planes); that is, the perpendicular ${2p_z}$ orbitals are easily affected by the Li-C chemical bonds. The critical differences among the stage-$n$ compounds could be utilized to evaluate the $N$-type doping effects.

Both Li$^+$-ion and Li-atom intercalations, as clearly shown in Figs. 10 and 9, exhibit the almost identical van Hove singularities in terms of their number and form, especially for the prominent shouder and peak structures due to the splitting $\pi$-electronic saddle points and the initial $\sigma$-electronic states. Compared with the Li$^+$ cases [Figs. 10(a)-10(d)], their corresponding energies in the stage-$n$ Li intercalation compounds present the red shift. The most important differences are the Fermi level and the 3D free carrier density. The Fermi level of the former is just located at the minimum density of states, being similar to the pristine system [Fig. 9(a)]. Apparently, this further illustrates that free electrons and holes are purely created by the interlayer atomic interactions of the C-${2p_z}$ orbitals. The Li$^+$ ions do not play critical roles on the carrier-related physical properties. However, the high-density conduction electrons in the Li-intercalation compounds are due to the significant charge transfer between Li and C atoms.

Scanning tunneling spectroscopy (STS) can efficiently examine the energy-dependent density of density states for valence and conduction bands in condensed-matter systems, according to the relation between current and bias voltage. A lot of I-V curves are performed by a probing tip being fixed at the same height and scanning on a surface. They are mainly determined by the surface electronic states, while the tunneling current is assumed to be roughly proportional to the constant states of a probing tip. The normalized differential conductance of the tip-surface tunneling junction is characterized as [dI/dV]/[I/V], being interpreted as the DOS of surface structure. As a result, the STS measurements are able to accurately identify the van Hove singularities due to the band-edge states and the semiconducting, semi-metallic and metallic behaviors. They have successfully verified the diverse electronic energy spectra in 1D graphene nanoribbons,\cite{huang2012spatially,sode2015electronic,chen2013tuning} 1D cylindrical carbon nanotubes,\cite{wilder1998electronic,odom1998atomic} 2D layered graphene systems,\cite{luican2011single,li2010observation,cherkez2015van,lauffer2008atomic,yankowitz2013local,que2015stacking,pierucci2015evidence} 2D adatom-adsorbed graphenes.\cite{chen2015long,chen2015tailoring,gyamfi2011fe} and 3D Bernal graphite,\cite{klusek1999investigations,li2009scanning} as measured from the energy, form, number, and intensity and form of special structures in DOSs. Concerning graphene nanoribbons, the edge- and width-dependent energy gaps and the square-root divergent peaks of 1D parabolic dispersions are confirmed from the precisely defined crystal structures\cite{magda2014room,huang2012spatially,sode2015electronic,chen2013tuning,wilder1998electronic,odom1998atomic} The prominent asymmetric peaks are also revealed in carbon nanotubes, clearly indicating the chirality- and radius-dominated band gaps and energy spacings between two neighboring subbands.\cite{wilder1998electronic,odom1998atomic} Specially, armchair nanotubes belong to the 1D metallic systems with a finite DOS at the Fermi level.\cite{wilder1998electronic} Many STS measurements on few-layer and adatom-graphenes show the low-lying DOS characteristics: a linear $E$-dependence vanishing at the Dirac point in monolayer graphene,\cite{li2009scanning} the asymmetry-induced peak structures in bilayer graphene,\cite{luican2011single,li2010observation,cherkez2015van} an electric-field-created band gap in bilayer AB stacking and tri-layer ABC stacking,\cite{lauffer2008atomic,yankowitz2013local} a prominent peak at $E_F$ related to the partially flat bands in tri-layer and penta-layer ABC stacking,\cite{wilder1998electronic,odom1998atomic} a dip structure at $E_F$ accompanied with a pair of asymmetric peaks in tri-layer AAB stacking,\cite{wilder1998electronic} and a red shift of Dirac point due to the metallic doping of bismuth adatoms.\cite{chen2015long,chen2015tailoring} The measured DOS of the AB-stacked graphite is finite near the Fermi level characteristic of the semi-metallic behavior\cite{li2009scanning} and displays the splitting $\pi$ and $\pi^\ast$ strong peaks at the deeper and higher energy, respectively.\cite{klusek1999investigations}

The high-resolution STS measurements can identify the diverse energy spectra of 3D graphite-related systems by examining the critical van Hove singularities. The main features of density states are very sensitive to the stacking configurations and the distribution and concentration of guest atoms/ions. For the pristine graphites, both AA and ABC stackings could be verified in the further STS experiments by measuring the DOS value at the Fermi level and  the widths of the prominent valence and conduction shoulder/peak structures covering ${+2}$ eV $\&$ $-2$ eV. Apparently, these two features are sufficiently in distinguishing three kinds of typical graphites, since they directly reflect the interlayer hopping integrals of the ${2p_z}$ orbitals.. The Fermi level, being relative to the minimum DOS, is shifted to the conduction-band states with a higher DOS after Li intercalation. The blue shift and its DOS value declines with the increment of $n$. Such examinations could identify the significant but weak Li-C chemical bondings. Foe the stage-$n$ Li graphite intercalation compounds, the experimental verifications on the $n$-split prominent peak/shoulder structures are very useful in understanding the chemical environments experienced by the  C-${2p_z}$. The similar phenomena are revealed in the stage-$n$ Li$^+$ systems, while their Fermi levels are almost identical to that of the pristine case. The latter results indicate the vanishing charge transfer between Li$^+$ ions and C atoms.

\vspace{5mm}
\par\noindent
{\bf Chemical bondings and charge distributions}

The spatial charge density and the variation of charge density, as clearly illustrated in Figs. 7(a)-7(i) and 8(a)-8(h), can provide very useful information about the chemical bondings and thus understand the dramatic transform of energy bands. The latter is created by subtracting the charge density of a simple hexagonal graphite and Li atoms/Li$^+$ ions from that of a graphite intercalation compound. ${\rho}$ and ${\Delta \rho}$, which are shown on the ${x-z}$ plane, are sufficiently in understanding the chemical bondings induced by the intercalation. Under the pristine case, there exists a very high charge density between two neighboring carbon atoms [any red region in Fig. 7(a)], obviously indicating the $\sigma$-covalent bondings due to to the planar ${(2p_x, 2p_y, 2s)}$ orbitals. Furthermore, the ${2p_z}$ orbitals perpendicular to the $z$-direction can form the wave-like $\pi$ bondings near the boundary of charge distribution. In addition, such bondings should have a weak but significant change to create the interlayer hopping integrals [Refs], compared with those in a monolayer graphene [Refs]. Their minor changes are also revealed among the AA-, AB- and ABC-stacked graphites [Refs].

After the Li-atom intercalations, the spatial charge distributions present the obvious changes, depending on the concentration and configuration of guest atoms. For the stage-1 compound, the range of the $\pi$-electronic charge distribution is extended between Li and C atoms, as observed from the shallow blue-color boundary [Fig. 7(b)]. This clearly indicate the orbital hybridizations in the Li-C bonds. Such range is redcued in the increment of stage $n$ owing to the declining number of Li-C bonds, e.g., ${\rho\,}$'s for the stage-2, stage-3 and stage-4 systems in Figs, 7(d), 7(f) and 7(h), respectively. Specifically, the charge density almost remains the same while the graphitic layers do not present the direct interactions with Li guest atoms, such as, ${\rho}$ on the internal one layer (two layers) in the stage-3 (stage-4) compound. The very strong evidence of ${2s-2p_z}$ hybridization in  Li-C bond is revealed in the charge variation. Apparently, the stage-1 system exhibits a quite obvious electron transfer from Li to C atoms, as shown by the heavy red region in fig. 7(c) covered by a black dashed rectangle. With the increasing $n$, ${\Delta \rho}$ in such range becomes small [the light red color in this region].

The Li$^+$-ion graphite intercalation compounds exhibit the unusual charge distribution, compared with the Li-atom cases. For the highest ion-concentration [Li$^+$C$_6$ in Fig. 8(a)], the low charge density comes to exist between Li$^+$ and C. The similar results are revealed in the nearest-neighboring graphitic layers of the stage-2, stage-3 and stage-4 systems [Figs. 8(c), 8(e), and 8(g). These are further illustrated by the charge-density variations after the Li$^+$ intercalations. A light charge redistribution appears at the co-dominating region of Li$+$ ions and C atoms, as obviously indicated by the shallow yellow regions in Figs. 8(b), (d), (f) and (h)]. Furthermore, it approaches to the former side, in which the effective charge transfer from C to Li$^+$ is estimated to be below 0.08 e per carbon atom. This phenomenon weakly depends on the stage number. The very low charge transfer might suggest the absence (or the weakness) of 2s-2p$_z$ hybridization in Li$^+$-C bond and the existence of the dipole-dipole interactions between the Li$^+$-ion and C-atom layers. On the other side, ${\Delta \rho}$'s in stage-3 and stage-4 cases related to the next-neighboring layers are almost absent [Figs. 8(f) and 8(h)], being similar to the  Li-atom cases [Figs. 7(g) and 7(i)].

\vspace{5mm}
\par\noindent
{\bf Summary}

Apparently, the 3D graphite-based systems exhibit the rich and unique essential properties. According to the delicate first-principles calculations, the typical AA-. AB- and ABC-stacked graphites have the ground states of ${-0.238}$ eV, ${-0.6}$ eV and ${-0.4}$ eV, respectively, corresponding to the optimal interlayer distances of 3.489 $\AA$, 3.35 $\AA$ and 3.40 $\AA$. The second and third stacking configurations consist of a natural graphite, especially for the dominating AB stacking. However. the AA stacking is revealed as the optimal geometric structure in the Li $\&$ Li$^+$ graphite intercalation compounds. The guest atoms/ions could form the extra planar structures.
The interlayer distances of two neighboring graphitic layers, with and without Li/Li$^+$ layers, are, respectively, enhanced and reduced after the chemical intercalation. The Li$^+$-ion intercalations create the shorter interlayer distances and have the lower ground state energies, compared with the Li-atom cases. Furthermore, the Fermi sea, which  could support the charge neutrality, might play a critical role in determining the total ground state energy. The theoretical predictions on the interlayer distances could be verified by the high-resolution TEM.

As to electronic energy spectra of graphite-related systems, AA, AB and ABC stackings, respectively, exhibit the linearly intersecting Dirac cones  at the corners (the first Brillouiz zone), the monolayer- $\&$ bilayer-like energy dispersions initiated from there, and the spiral Dirac-cone structure near the corners. The $k_z$-dependent energy widths are, respectively, ${\sim\,1.0}$ eV, ${\sim\,0.2}$ eV and ${0.02}$ eV. Apparently, the AA-stacked (ABC-stacked) possesses the largest (smallest) overlap of valence and conduction bands and thus the highest (lowest) free electron and hole densities, indicating the most effective  cooperations (the strong competitions) of the interlayer hopping integrals.. There are more low-lying energy subbands after the atom/ion intercalation, mainly owing to the enlarged unit cell. An obvious blue shift of the Fermi level is only revealed in the stage-$n$ Li graphite intercalation compounds, but not the Li$^+$ ones. This quantity declines in the increment of $n$, clearly illustrating the reduced total charge transfer from Li guest atoms to C host ones. The high-density conduction electrons are induced by the Li-atom intercalation, while only the low-density free electrons  and holes, respectively, in conduction and valence bands survive under the Li$^+$-ion cases. Such free carriers are, respectively, induced by the charge transfer in the Li-C bonds and the interlayer hopping integrals of the C-${2p_z}$ orbitals. Another  focus is the $\pi$-electronic valence energy spectrum, corresponding to parabolic dispersions with the saddle points on the middles of the first Brillouin zone. The stage-$n$ Li/Li$^+$ graphite intercalation compounds exhibit the $n$-split $\pi$-electronic (${n/2}$-split $\sigma$-electronic) valence subbands below the Fermi level more than 1 (3) eV. This further illustrates the fact that the distribution and concentration of guest atoms/ions can create the diverse chemical environments experienced by the $\pi$ and $\sigma$ electrons.

The main features of energy bands are revealed as the critical van Hove singularities in density of states. For the $\pi$-electronic energy spectra in pristine graphites, the AA (ABC) possesses the largest (smallest) at the Fermi level and the widest (narrowest) shoulder structures at middle energies, Specifically, their most prominenet contributions might merge with the initial $\sigma$-electronic ones. Such association is split by the intercalation effect. The stage$-n$ atom-intercalated graphite compounds present the blue-shift Fermi level with a higher DOS at the conduction energy spectrum; furthermore, there exist the $n$-split (${n/2}$-split) peak/shoulder structures for the valence and  conduction $\pi$-electronic spectrum (the valence         $\sigma$-electronic spectrum). The splitting behaviors also appear in the stage-$n$ Li$^+$-ion cases, while the Fermi levels correspond to the minimum DOSs. The high-resolution STS measurements on these two critical characteristics can distinguish the distinct charge transfers and chemical environments.

Apparently, the spatial charge distributions of graphite-related systems exhibit the rich and unique features. A simple hexagonl graphite has the very strong covalent $\sigma$ bondings due to the ${2p_x,2p_y,2s}$ orbitals on the graphitic planes, being hardly affected by the guest atoms/ions. Furthermore, the weaker $\pi$ bondings of ${2p_z}$ orbitals are extended on the ${(x,y)}$-planes, so they will dominate the low-energy essential properties. The distribution ranges of $\pi$ bondings are widened after the Li and Li$^+$ intercalations. The very obvious charge redistributions are revealed between Li and C atoms, in which the electron charge transfer from the former to the later is estimated to be about 0.85 from the Bader analyses. These results clearly indicate that the significant Li-C bonds are created by the non-negligible ${2s-2p_z}$ orbital hybridizations, and they lead to the high-density of free conduction electrons, as identified from the large blue shift of the Fermi level. On the other hand, only the light charge-density variations come to exist between Li$+$ ions and C atoms, and the charge transfer from the latter to the former is roughly 10$\%$ of the atom-intercalation cases. The dominating intercations of host atoms and guest ions might belong to the dipole-dipole forms, but not the orbital-orbital ones.

\newpage
${\centerline {\bf Figure Captions}}$

Figure 1: The flow chart of VASP calculations

Figure 2: The dependence of the total ground state energy on the interlayer distance for
(a) a simple graphite, (b) LiC$_6$ and (c) Li$^+$C$_6$.

Figure 3: Geometric structures: (a) a simple hexagonal graphite, (b) top views of a pristine system and LiC$_6$,
(c) LiC$_6$/Li$^+$C$_6$ (stage-1 lithium graphite intercalation compound), (d) LiC$_{12}$/Li$^+$C$_{12}$ (stage-2), (e) LiC$_{18}$/Li$^+$C$_{18}$ (stage-3), and (f) LiC$_{24}$/Li$^+$C$_{24}$ (stage-4).

Figure 4: Electronic structures for the (a) AA-, (b) AB-stacked pristine graphites, and (c) ABC-stacked graphites. Also shown in (d) is that of the first system under the enlarged unit cell with six carbon atoms, and (e) correspond to the original and reduced first Brilloun zones.

Figure 5: Valence and conduction bands of (a) LiC$_6$, (b) LiC$_{12}$, (c) LiC$_{18}$, and (d) LiC$_{24}$.

Figure 6: Similar plots as Figs. 5(a)-5(d), but shown for (a) Li$^+$C$_6$, (b) Li$^+$C$_{12}$, (c) Li$^+$C$_{18}$, and (d) Li$^+$C$_{24}$.

Figure 7: The spatial charge distributions before/after Li intercalation: (a) a simple hexagonal graphite,
(b)/(c) LiC$_6$, (d)/(e) LiC$_{12}$, (f)/(g) LiC$_{18}$, and (h)/(i) LiC$_{24}$.

Figure 8: Similar plots as Figs. 5(b)-5(i), but indicated for (a)/(b) Li$^+$C$_6$, (c)/(d) Li$^+$C$_{12}$, (e)/(f) Li$^+$C$_{18}$, and (g)/(h) Li$^+$C$_{24}$.

Figure 9: The carbon-orbital-projected DOSs: (a) the AA-, (b) AB- and (c) ABC-stacked graphites.

Figure 10: The C- and Li-orbital-decomposed DOSs for (a) LiC$_6$, (b) LiC$_{12}$, (c) LiC$_{18}$ (a) LiC$_{24}$.

Figure 11: Similar DOSs as Figs. 10(c)-10(f), but displayed for (a) Li$^+$C$_6$, (b) Li$^+$C$_{12}$, (c) Li$^+$C$_{18}$, and (d) Li$^+$C$_{24}$.

\begin{figure}[htbp]\centering
\includegraphics[width=0.9\linewidth]{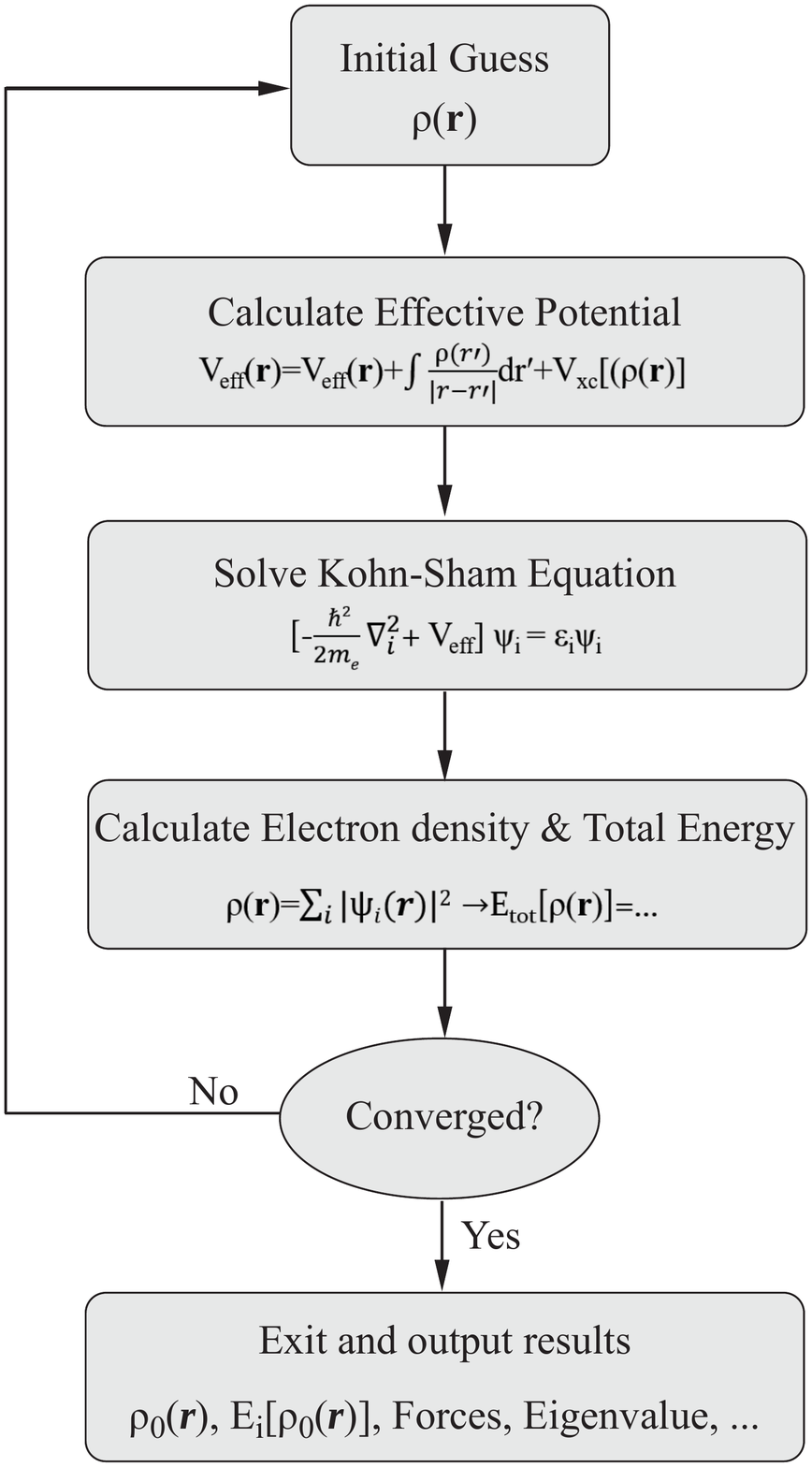}
\caption{The flow chart of VASP calculations}
\end{figure}

\begin{figure}[htbp]\centering
\includegraphics[width=0.9\linewidth]{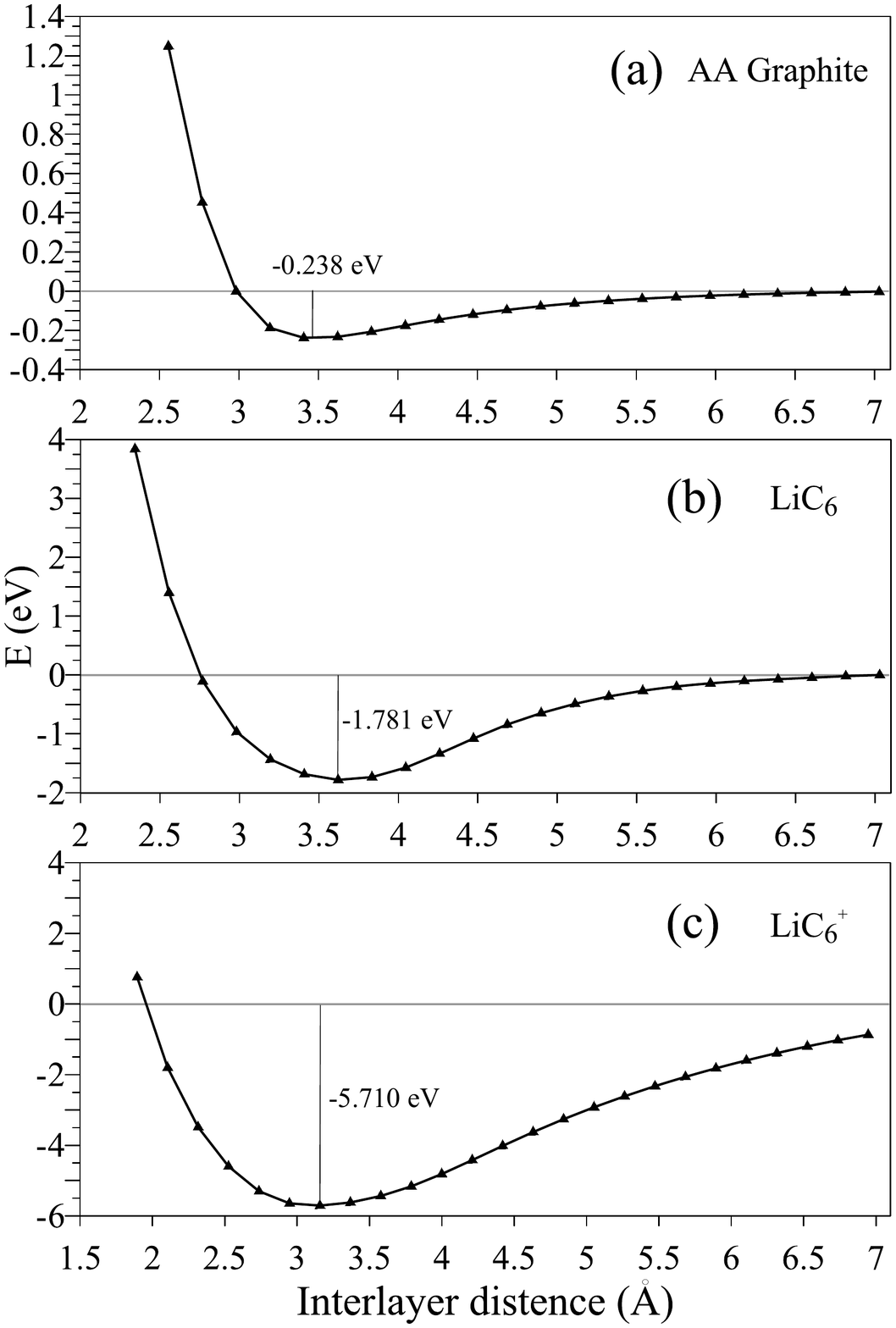}
\caption{The dependence of the total ground state energy on the interlayer distance for
(a) a simple graphite, (b) LiC$_6$ and (c) Li$^+$C$_6$.}
\end{figure}

\begin{figure}[htbp]\centering
\includegraphics[width=0.9\linewidth]{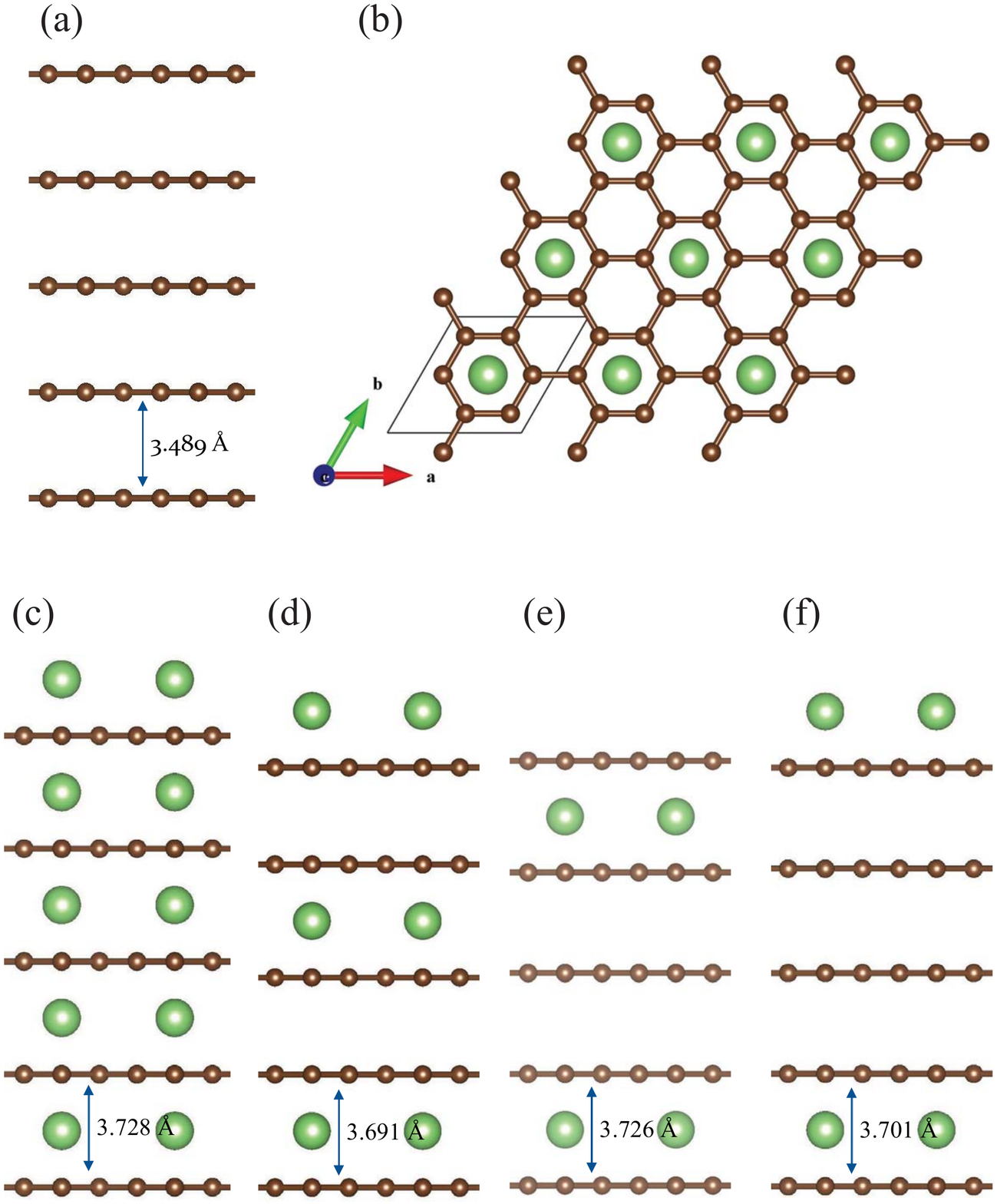}
\caption{Geometric structures: (a) a simple hexagonal graphite, (b) top views of a pristine system and LiC$_6$,
(c) LiC$_6$/Li$^+$C$_6$ (stage-1 lithium graphite intercalation compound), (d) LiC$_{12}$/Li$^+$C$_{12}$ (stage-2), (e) LiC$_{18}$/Li$^+$C$_{18}$ (stage-3), and (f) LiC$_{24}$/Li$^+$C$_{24}$ (stage-4).}
\end{figure}

\begin{figure}[htbp]\centering
\includegraphics[width=0.9\linewidth]{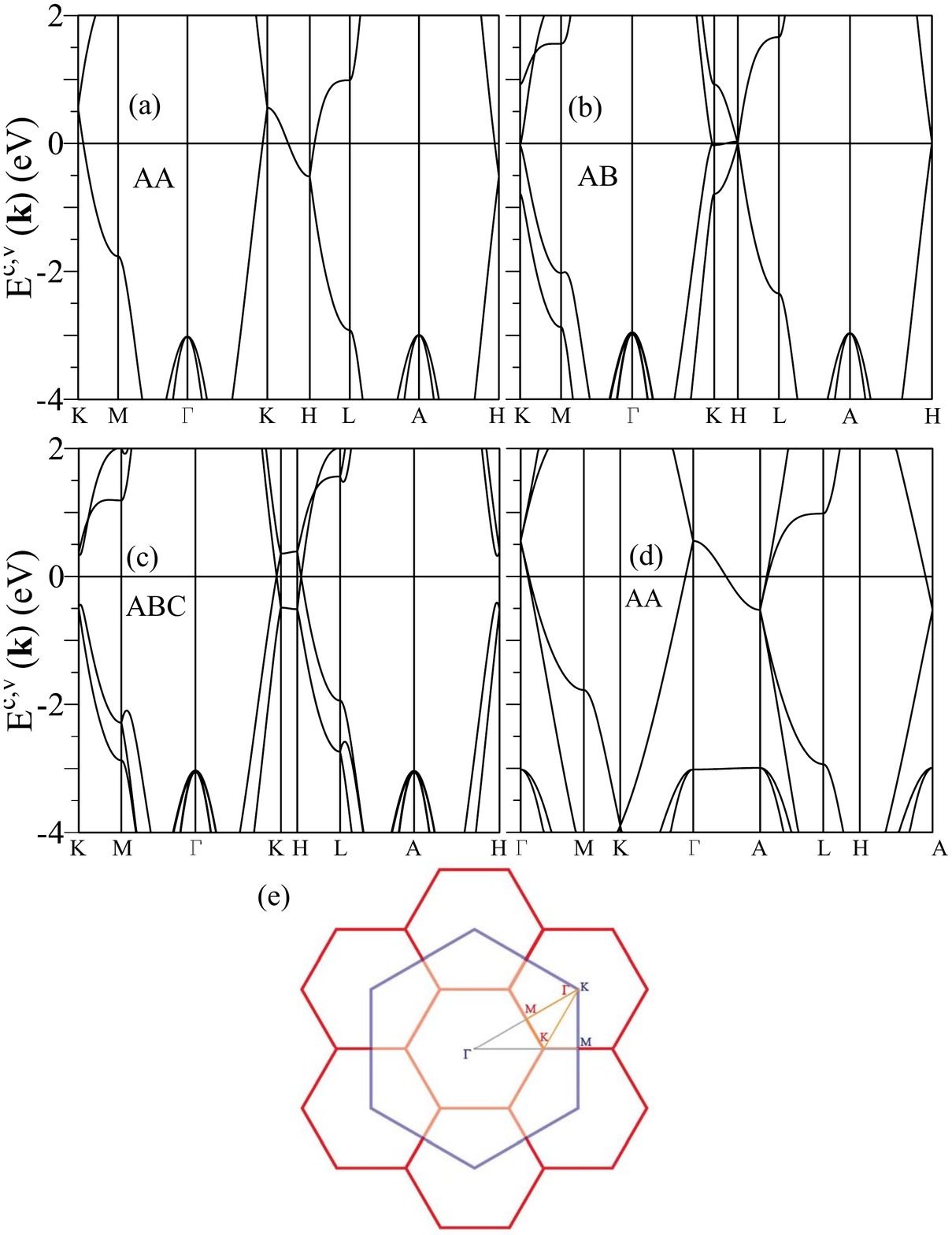}
\caption{Electronic structures for the (a) AA-, (b) AB-stacked pristine graphites, and (c) ABC-stacked graphites. Also shown in (d) is that of the first system under the enlarged unit cell with six carbon atoms, and (e) correspond to the original and reduced first Brilloun zones.}
\end{figure}

\begin{figure}[htbp]\centering
\includegraphics[width=0.9\linewidth]{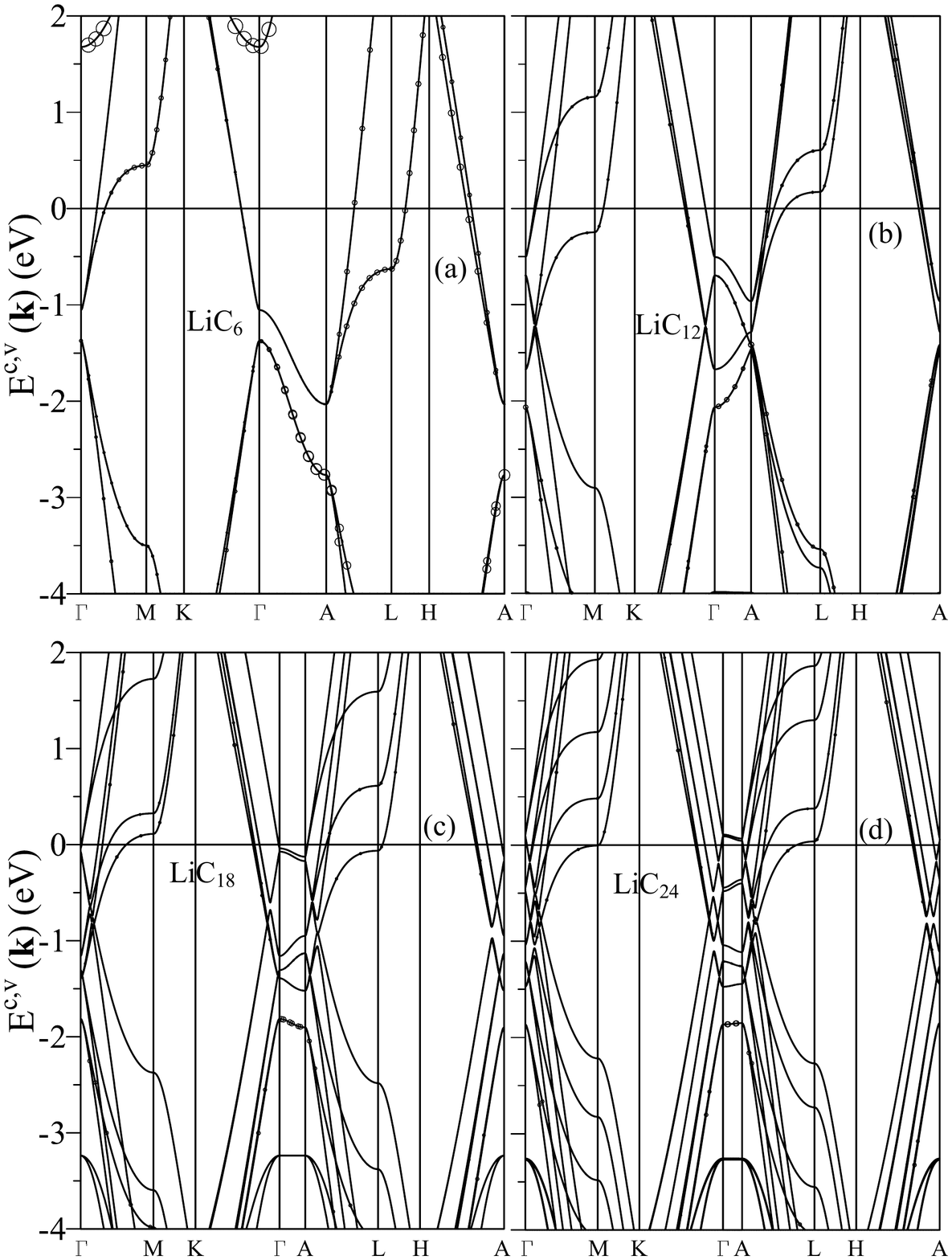}
\caption{Valence and conduction bands of (a) LiC$_6$, (b) LiC$_{12}$, (c) LiC$_{18}$, and (d) LiC$_{24}$.}
\end{figure}

\begin{figure}[htbp]\centering
\includegraphics[width=0.9\linewidth]{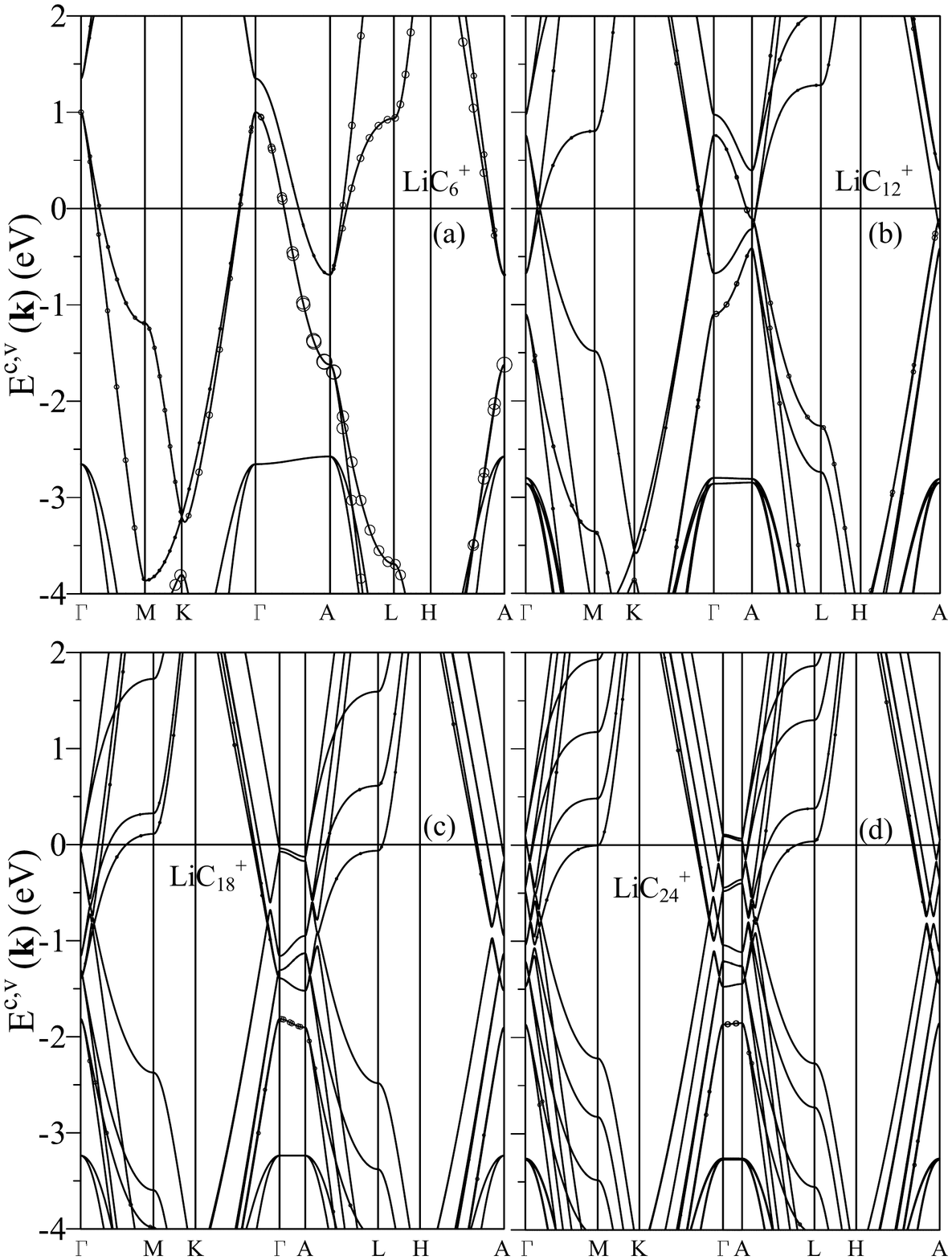}
\caption{Similar plots as Figs. 5(a)-5(d), but shown for (a) Li$^+$C$_6$, (b) Li$^+$C$_{12}$, (c) Li$^+$C$_{18}$, and (d) Li$^+$C$_{24}$.}
\end{figure}

\begin{figure}[htbp]\centering
\includegraphics[width=0.9\linewidth]{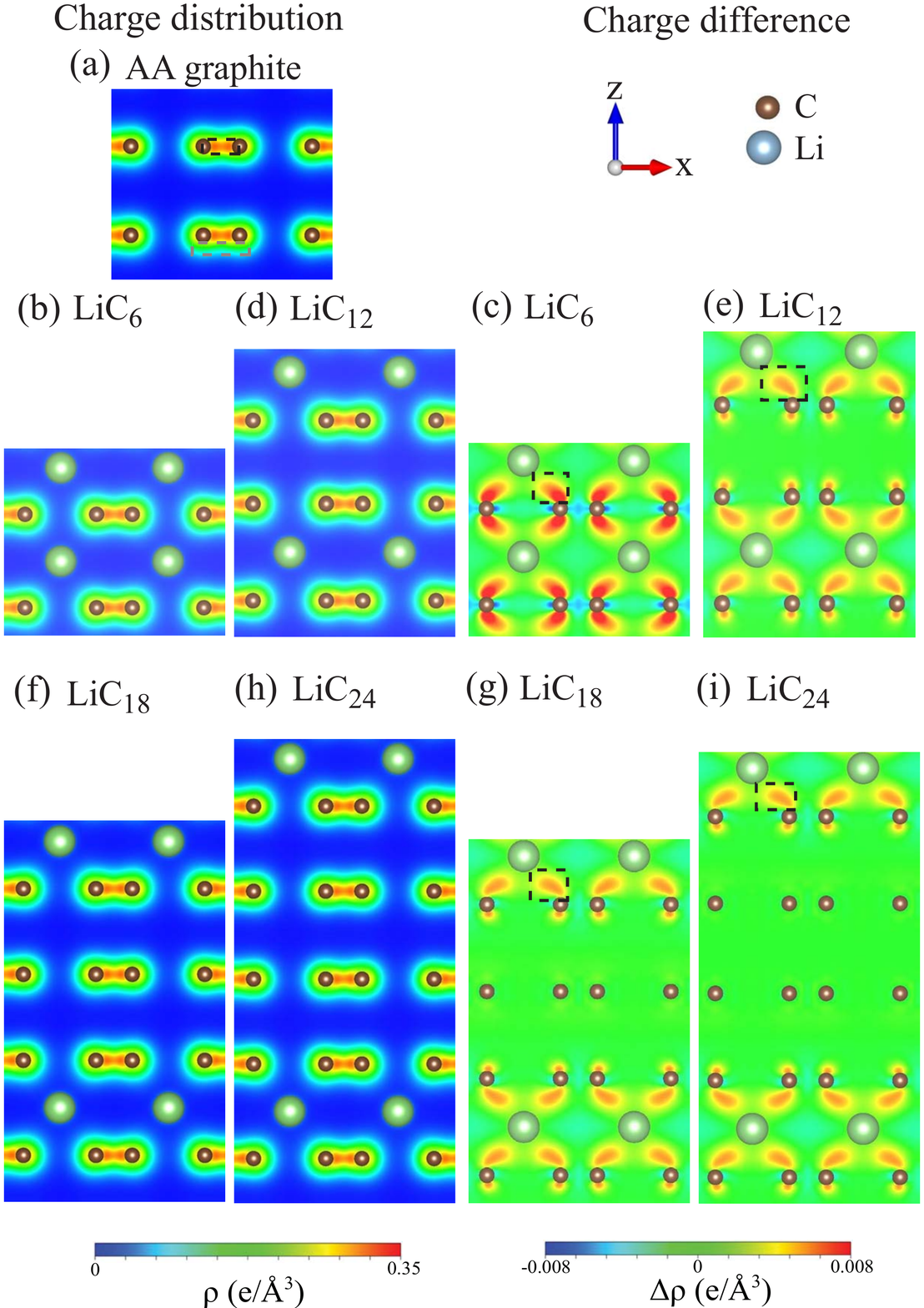}
\caption{The spatial charge distributions before/after Li intercalation: (a) a simple hexagonal graphite,
(b)/(c) LiC$_6$, (d)/(e) LiC$_{12}$, (f)/(g) LiC$_{18}$, and (h)/(i) LiC$_{24}$.}
\end{figure}

\begin{figure}[htbp]\centering
\includegraphics[width=0.9\linewidth]{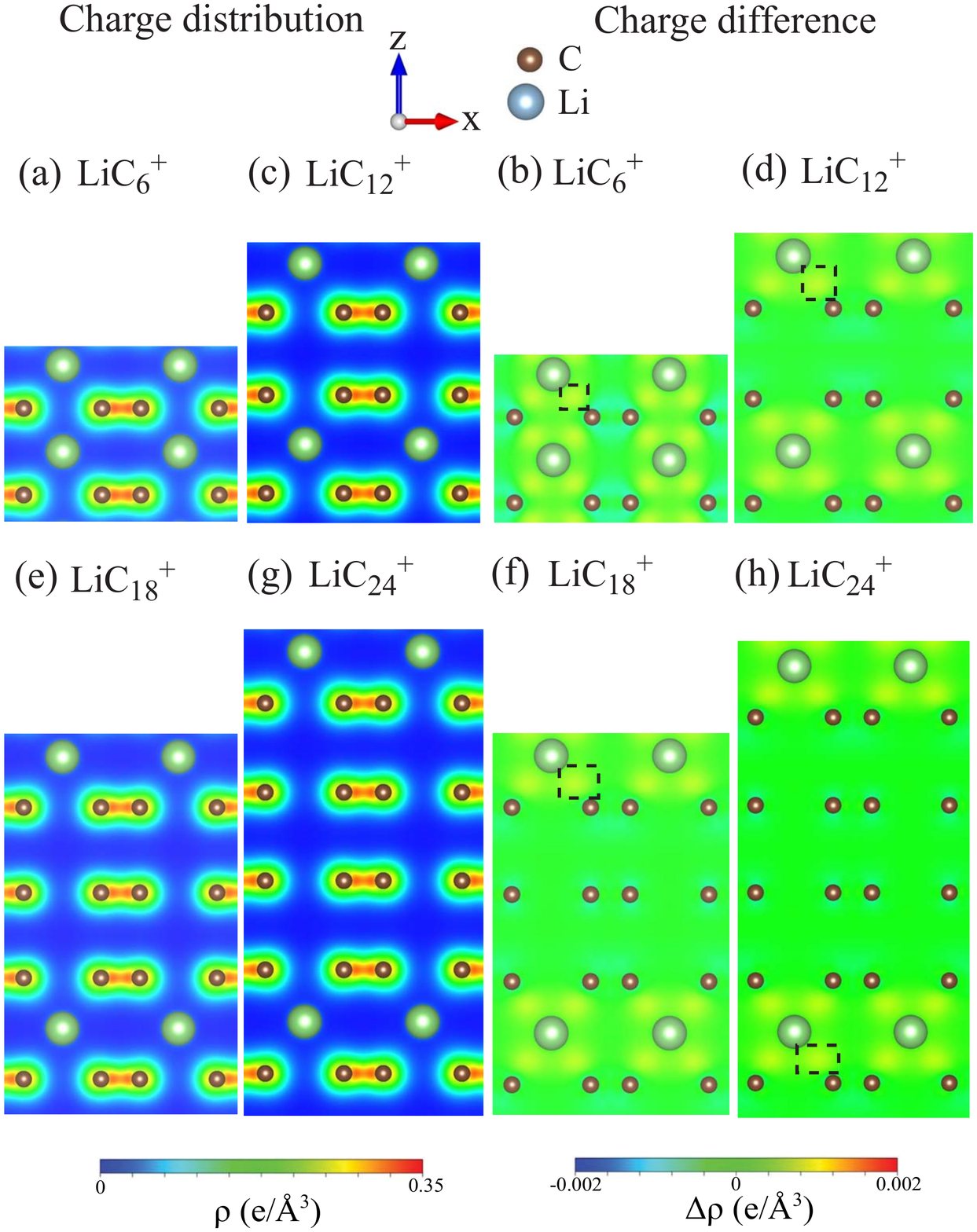}
\caption{Similar plots as Figs. 5(b)-5(i), but indicated for (a)/(b) Li$^+$C$_6$, (c)/(d) Li$^+$C$_{12}$, (e)/(f) Li$^+$C$_{18}$, and (g)/(h) Li$^+$C$_{24}$.}
\end{figure}

\newpage
\bibliography{achemso}

\end{document}